\documentclass{article}
\usepackage{natbib,epsfig,multirow}

\usepackage{amsfonts}
\usepackage{natbib}
\usepackage{colonequals}
\usepackage{amsfonts} 
\usepackage{amsmath}
\usepackage{subfigure}
\usepackage{amssymb,lineno}
\usepackage[usenames,dvipsnames]{xcolor}
\usepackage{mathrsfs}
\usepackage{booktabs}
\usepackage{graphicx}
\usepackage{subfig}
\usepackage{multirow}
\usepackage{array}

\newcommand{\ident}{\mathbf{I}}

\newcommand{\vecx}{\mathbf{x}}
\newcommand{\vecX}{\mathbf{X}}

\newcommand{\mata}{\mathbf{A}}
\newcommand{\matd}{\mathbf{D}}

\newcommand{\varthet}{\mbox{\boldmath$\vartheta$}}
\newcommand{\vectheta}{\mbox{\boldmath$\theta$}}
\newcommand{\vecTheta}{\mbox{\boldmath$\Theta$}}
\newcommand{\vecmu}{\mbox{\boldmath$\mu$}}

\newcommand{\mSigma}{\mbox{\boldmath$\Sigma$}}
\newcommand{\matsig}{\mSigma}

\begin{document}

\title{Mixture Model Averaging for Clustering}
\author{Yuhong Wei\thanks{Department of Mathematics \& Statistics, University of Guelph, Guelph, Ontario, Canada, N1G~2W1.} and Paul~D.~McNicholas\thanks{Department of Mathematics \& Statistics, McMaster University, Hamilton, Ontario, Canada,  L8S~4L8. Tel.: 905-525-9140, ext.\ 23419. Email {\tt mcnicholas@math.mcmaster.ca}}}
\date{}
\maketitle

\begin{abstract}
In mixture model-based clustering applications, it is common to fit several models from a family and report clustering results from only the `best' one. In such circumstances, selection of this best model is achieved using a model selection criterion, most often the Bayesian information criterion. Rather than throw away all but the best model, we average multiple models that are in some sense close to the best one, thereby producing a weighted average of clustering results. Two (weighted) averaging approaches are considered: averaging the component membership probabilities and averaging  models. In both cases, Occam's window is used to determine closeness to the best model and weights are computed within a Bayesian model averaging paradigm. In some cases, we need to merge components before averaging; we introduce a method for merging mixture components based on the adjusted Rand index. The effectiveness of our model-based clustering averaging approaches is illustrated using a family of Gaussian mixture models on real and simulated data. 
\end{abstract}

\section{Introduction}\label{sec:intro}
Model averaging approaches provide a coherent mechanism for accounting for model uncertainty through combining parameter estimates across a set of competing models. 
Bayesian model averaging \citep[BMA; cf.][]{madigan94,raftery95} is a popular approach to model averaging that has been applied successfully to many statistical models, including linear regression \citep{raftery98}, generalized linear models \citep{raftery96}, Cox regression models \citep*{hoeting99a}, and survival analysis \citep{volinsky97}. A comprehensive review of Bayesian model averaging is given by \cite*{hoeting99}. Despite its prevalence, BMA has not previously been applied in clustering applications, where there may be different clusterings for the same data set emanating from different models. 

While BMA has not been used, other approaches for combining several clustering results to give a single clustering have been tried. One such method, based on least-squares, was proposed by \cite{dahl06} and has been applied to Bayesian mixture models \citep[e.g.,][]{molitor10}. 
Another class of methods, called consensus clustering, has also been proposed. \cite{fred05} summarize various clusterings in a co-association matrix, which indicates the strength of association between objects by analyzing how often each pair of objects appears in the same cluster. The final clustering is then determined using a voting-type algorithm. \cite{strehl02} propose three graph-based consensus clustering algorithms: the cluster-based similarity partitioning algorithm (CSPA), the hyper-graph partitioning algorithm (HGPA), and the meta-clustering algorithm (MCLA). A problem arises when the various clusterings have different numbers of clusters, and the same problem arises in our BMA approach; we resolve it by merging mixture components (Section~\ref{sec:merging}).

Model-based clustering is an idiom that is often used to describe the application of a mixture model for clustering. Dating at least as far back as \citet{wolfe63}, interest in model-based clustering is increasing steadily in application areas such as food authenticity, social networks, and microarray gene expression analyses \citep[e.g.,][]{yeung01,wehrens04,krivitsky09,mcnicholas10}. In model-based clustering applications, it is common to fit many mixture models within a family (cf.\ Section~\ref{sec:back}) and then report clustering results based only on the best one. Such criterion-based methods of model selection have the general feature that the smaller (or larger, as the case may be) the value of the criterion, the stronger the `evidence' for the model (i.e., the covariance structure and number of components, as well as number of latent variables if relevant). 

In this paper, we argue that criterion-based approaches that throw away all but the best model are not necessarily optimal. One may argue, inferentially, that it should matter that many models have been fitted when reporting clustering results. 
Furthermore, criterion-based approaches may be unreliable when the difference between the best value of a criterion is close to one or more other values. The goal of this paper is to depart from the paradigm 
of selecting a single best model within the mixture model family setting.
Instead, we report clustering results based on an averaging of models. While model averaging in itself is not a novel idea, its use within the model-based clustering literature marks a significant departure from the accepted norm. We consider two different approaches to averaging herein: averaging models to produce one interpretable model and directly averaging \textit{a~posteriori} probabilities.

The remainder of this paper is outlined as follows. In Section~\ref{sec:back}, background on model-based clustering is provided. In Section~\ref{sec:method}, we describe our model averaging schemata, including a novel approach for merging components, before illustrating our approaches on real data (Section~\ref{sec:data}). The paper concludes with discussion and suggestions for future work (Section~\ref{sec:conc}).

\section{Model-Based Clustering}\label{sec:back}
\subsection{Gaussian Parsimonious Clustering Models}
A $p$-dimensional random vector $\vecX$ arises from a parametric finite mixture distribution if its density can be written $f(\vecx~|~\varthet)=\sum_{g=1}^G\pi_g\, f_g(\vecx~|~\vectheta_g)$,  where $G$ is the number of components, $\pi_g$ are mixing proportions $(\pi_g>0,\sum_{g=1}^G\pi_g=1)$, and $\varthet$ denotes the model parameters, i.e., $\varthet=(\pi_1, \ldots, \pi_G, \vectheta_1,\ldots, \vectheta_G)$. 
Until very recently, Gaussian mixtures have dominated the model-based clustering literature. The likelihood for $p$-dimensional $\vecx_1,\ldots,\vecx_n$ from a Gaussian mixture model is 
\begin{equation}\label{eqn:hmm}
\mathcal{L}(\varthet)=\prod_{i=1}^n\sum_{g=1}^G\pi_g\phi(\vecx_i~|~\vecmu_g,\matsig_g),
\end{equation} 
where $\phi(\vecx_i~|~\vecmu_g,\matsig_g)$ is the density of a $p$-dimensional multivariate Gaussian distribution with mean $\vecmu_g$ and covariance matrix $\matsig_g$, and $\varthet$ once again denotes the model parameters. 

The density in \eqref{eqn:hmm} has a total of $(G-1)+Gp+Gp(p+1)/2$ free parameters and so it is usually necessary to introduce parsimony. A total of $Gp(p+1)/2$ of these parameters are in the component covariances, so imposing the isotropic constraint on the component covariances, i.e., $\matsig_g=\sigma_g\ident_p$, is a very simple way to reduce the number of parameters from quadratic to linear in~$p$. Of course, this constraint --- implying spherical components with different volumes --- will not be practical for most applications and so less restrictive constraints are needed. Several such approaches have been tried, usually based on imposing constraints upon a decomposed component covariance structure. The most famous such approach is based on an eigen-decomposition of the component covariance matrices \citep{banfield93}, i.e., $\matsig_g=\lambda_g\matd_g\mata_g\matd_g'$, where $\matd_g$ is the orthogonal matrix of eigenvectors of $\matsig_g$, $\mata_g$ is a diagonal matrix, with $|\mata_g|=1$, containing elements proportional to the eigenvalues of $\matsig_g$, and $\lambda_g$ is the associated constant. \cite{celeux95} impose constraints on $\lambda_g$, $\matd_g$, and $\mata_g$ to obtain a family of 14 Gaussian parsimonious clustering models (GPCMs; \tablename~\ref{models}). The ${\tt mixture}$ package  \citep{browne13} for the {\sf R} software \citep{R12} provides an implementation of all 14 members of the GPCM family. Parameter estimation for 12 of the members of the GPCM family is carried out using the expectation-maximization (EM) algorithm \citep{dempster77} and details are given by \cite{celeux95}. An MM algorithm \citep[cf.][]{hunter04} is used for the other two models (EVE and VEE), and extensive details are provided by \cite{browne13b}.
\begin{table}[ht]{\small
	\caption{Nomenclature, covariance structure, and number of covariance parameters for each member of the GPCM family.}
	\label{models}
	\centering{\small
	\begin{tabular}{l|lllrr}
	\hline
	 & {Volume} & {Shape} & {Orientation} & $\matsig_g$ & {Free cov.\ paras.}\\
	\hline
	EII &  Equal & Equal & NA & $\lambda \textbf{I}$ & 1\\
	VII & Variable & Equal & NA & $\lambda_g \textbf{I}$ & $G$\\
	EEI & Equal & Equal & Cord.\ Axes & $\lambda \textbf{A}$ & $p$\\
	VEI & Variable & Equal & Cord.\ Axes & $\lambda_g \textbf{A}$ & $p+G-1$ \\
	EVI & Equal & Variable & Cord.\ Axes & $\lambda \textbf{A}_g$ & $Gp-G+1$\\
	VVI & Variable & Variable & Cord.\ Axes & $\lambda_g \textbf{A}_g$ & $Gp$ \\
	EEE & Equal & Equal & Equal & $\lambda \textbf{D} \textbf{A} \textbf{D}'$ & $p(p+1)/2$\\
	EEV & Equal & Equal & Variable & $\lambda \textbf{D}_g \textbf{A} \textbf{D}_{g}'$ & $Gp(p+1)/2-(G-1)p$\\
	VEV & Variable & Equal & Variable & $\lambda_g \textbf{D}_g \textbf{A} \textbf{D}_{g}'$ & $Gp(p+1)/2-(G-1)(p-1)$\\
	VVV & Variable & Variable & Variable & $\lambda_g \textbf{D}_g \textbf{A}_g \textbf{D}_{g}'$ & $Gp(p+1)/2$\\
	EVE & Equal & Variable & Equal & $\lambda \textbf{D} \textbf{A}_g \textbf{D}'$ & $p(p+1)/2+(G-1)(p-1)$\\
	VVE & Variable & Variable & Equal & $\lambda_g \textbf{D} \textbf{A}_g \textbf{D}'$ & $p(p+1)/2+(G-1)p$\\
	VEE & Variable & Equal & Equal & $\lambda_g \textbf{D} \textbf{A} \textbf{D}'$ & $p(p+1)/2+(G-1)$\\
	EVV & Equal & Variable & Variable & $\lambda \textbf{D}_g \textbf{A}_g \textbf{D}_{g}'$ & $Gp(p+1)-(G-1)$\\	
	\hline
	\end{tabular}}}
\end{table}

A typical application of the GPCM family of models consists of running each of the models (\tablename~\ref{models}) for a range of values of $G$. Then, the best of these models is selected using some criterion and the associated classifications are reported. The most popular criterion for this purpose is the Bayesian information criterion \citep[BIC;][]{schwarz78}, i.e.,
$$\text{BIC} = -2l( \textbf{x}, \hat{ \boldsymbol{\vartheta} }) + \rho\log n,$$ 
where $\hat{\varthet}$ is the maximum likelihood estimate of $\varthet$, $l(\textbf{x},\hat{\varthet})$ is the maximized log-likelihood, and $\rho$ is the number of free parameters. \citet{leroux92} and \citet{keribin00} give theoretical results that, under certain regularity conditions, support the use of the BIC for choosing the number of components in a mixture model. In addition, model selection has been based on the BIC in a wide range of model-based clustering applications (recent examples include \citealp{bouveyron07}, \citealp{mcnicholas08}, \citealp{andrews11a}, and \citealp{vrbik14}). Nonetheless, the model with the smallest BIC does not necessarily give the best predicted classifications or the `correct' number of components \citep[see][for discussion]{biernacki00}.

\subsection{\textit{A Posteriori} Merging of Components}\label{sec:apmc}

In Gaussian model-based clustering, clusters are often taken as synonymous with mixture components. However, a cluster might itself be well modelled by a mixture of Gaussian distributions. In such cases, mixture components can be \textit{a~posteriori} merged to obtain clusters, and different approaches have been considered to do this, e.g., \cite{baudry10} and \cite{hennig10}.  \cite{baudry10} proposed finding the number of clusters with a hierarchical merging procedure that minimizes the entropy of the clustering solution obtained by fitting the MCLUST family to the data --- the MCLUST family is a subset of 10 of the GPCM models \citep[cf.][]{fraley02a}. \cite{hennig10} followed a similar hierarchical merging procedure, where some aggregation criteria are proposed based on either modality (i.e., merging components that produce a unimodal distribution) or on misclassification probabilities (i.e., merging components to minimize misclassification probabilities). In our merging procedure, we maximize the adjusted Rand index \citep[ARI;][]{hubert85} with a reference model. 

Recent work using non-Gaussian mixture models for clustering has brought into sharp relief the fact that, although it can work very well in some cases, merging Gaussian components is not a ``get out of jail free card" \citep{mcnicholas13}. The method of \cite{baudry10}, and accompanying software, is discussed by \cite{vrbik14}, who compare the \textit{a~posteriori} merging approach of \cite{baudry10} with results from fitting mixtures of skew-normal distributions and mixtures of skew-$t$ distributions. In these experiments, the merging approach gave inferior results to fitting mixtures with more flexible component densities. Similar results around the inferiority of merging components over fitting non-Gaussian mixtures have been reported by others, including \cite{murray13} and \cite{franczak14}. While not considered herein, our averaging approaches could be used to average non-Gaussian mixtures or \textit{a~posteriori} probabilities arising therefrom. Of course, great care would need to be taken when averaging mixtures of skewed distributions, especially in cases where each component has its own skewness parameter.

\subsection{Bayesian Model Averaging}\label{sec:bma}

Statistical inference with families of mixture models typically involves selection of a model using some criteria and then proceeding as if the selected model has generated the data, without considering the additional uncertainty introduced by ignoring all other models. 
Difficulties arise with this approach because it is practically impossible to decide what to do when the difference between the values of the criteria for two different models is `small'. Model averaging, including frequentist model averaging \citep[e.g.,][]{hjort03} and BMA, takes model uncertainty into consideration by combining parameter estimates across different models. Bayesian model averaging is a popular technique for model averaging, cf.\ \cite*{hoeting99}. Borrowing their notation, suppose we have models $\mathcal{M}_1, \mathcal{M}_2,\ldots,\mathcal{M}_K$, and $\Delta$ is the quantity of interest. The posterior distribution of $\Delta$ given data $D$ is
\begin{equation}
	\label{eq:triangleup}
  	\text{pr} (\Delta \mid D ) = \sum_{i=1}^{K} \text{pr} (\Delta \mid \mathcal{M}_i ,D ) \text{pr} ( \mathcal{M}_i \mid D ),
\end{equation}
where $\text{pr} (\Delta \mid \mathcal{M}_i ,D)$ is the posterior distribution of $\Delta$ under model $\mathcal{M}_i$, $\text{pr} ( \mathcal{M}_i \mid D )$ is the posterior probability for model $\mathcal{M}_i$, i.e.,
\begin{equation}
	\label{eq:postmodel}
	\text{pr} ( \mathcal{M}_i \mid D)= \frac{\text{pr}(D \mid \mathcal{M}_i) \text{pr}(\mathcal{M}_i)}{\sum_{i=1}^K \text{pr}(D \mid \mathcal{M}_i) \text{pr}(\mathcal{M}_i)},
\end{equation}
and 
\begin{equation}
	\label{eq:integratedlike}
	\text{pr}(D \mid \mathcal{M}_i) = \int_{\vecTheta_i} {\text{pr}(D \mid \vectheta_i , \mathcal{M}_i) \text{pr}(\vectheta_i \mid \mathcal{M}_{i}) d{\vectheta_i}},
\end{equation}
where 
$\vectheta_i$ is the vector of parameters of model $\mathcal{M}_i$, $\text{pr}(\vectheta_i \mid \mathcal{M}_i)$ is the prior for~$\vectheta_i$ under model $\mathcal{M}_i$, and $\text{pr}(\mathcal{M}_i)$ is the prior probability of model $\mathcal{M}_i$. 

\cite{madigan94} show that BMA can give better predictive performance than any single model; however, BMA 
has two major implementation difficulties. One is that the number of models in the summation in \eqref{eq:triangleup} can be prohibitively large. Another is that the posterior model probabilities are hard to compute because they involve very high-dimensional integrals. To 
overcome the former problem, \cite{madigan94} proposed using Occam's window to choose a set of models. The idea is that if a given model is a far worse fit to the data than the best model, then it is discarded and no longer considered. More formally, models not in Occam's window 
\begin{equation}
	\label{eq:owindow}
\left \{  \mathcal{M}_i :  \frac{\max_{l} \left \{ \text{pr}(\mathcal{M}_l \mid D)\right \}} {\text{pr}(\mathcal{M}_i \mid D)} \leq c  \right \}
\end{equation} are discarded, where $c$ is some positive number. \cite{madigan94} use $c=20$ in their analyses, drawing an analogy to a $p$-value of $0.05$, and we use the same value in our analyses to determine which models to average (Section~\ref{sec:data}). 

In the case of mixture models, the BIC can be used to approximate the integral in \eqref{eq:integratedlike} \citep{kass95,dasgupta98}. Specifically, the BIC approximation can be used to compute $\text{pr}(D \mid \mathcal{M}_i)$ via 
$$\text{pr}(D \mid \mathcal{M}_i) = \exp\left\{-\frac{1}{2}\text{BIC}_i\right\},$$ 
where $\text{BIC}_i$ is the BIC value for model $\mathcal{M}_i$.
Therefore, under equal prior probabilities, $\text{pr}(\mathcal{M}_i \mid D)$ can be computed based on \eqref{eq:postmodel}, i.e.,
\begin{equation}
	\label{eq:postmodel2}
	\text{pr} ( \mathcal{M}_i \mid D)= \frac{\exp\left\{-\frac{1}{2}\text{BIC}_i\right\}}{\sum_{i=1}^K \exp\left\{-\frac{1}{2}\text{BIC}_k\right\}},
\end{equation}
and Occam's window \eqref{eq:owindow} is equivalent to 
\begin{equation}\label{eqn:window}
\big\{\mathcal{M}_i : \text{BIC}_i - \min_l\{\text{BIC}_l\} \leq 2\log c\big\}.
\end{equation}

Herein, we use \eqref{eq:postmodel2} to compute the weights for our averaging approaches and we use Occam's window \eqref{eqn:window} to decide which models to average. Before describing our averaging approaches, we need to consider how to merge mixture components; this will be necessary when models in Occam's window have different numbers of components and we want to use all of the models.

\section{Methodology}\label{sec:method}

\subsection{Merging Mixture Components}\label{sec:merging}
In this section, we introduce a procedure for merging mixture components. Suppose we have a $G$-component mixture model and we want to merge components to produce an $H$-component mixture, where $H < G$. The density of the resulting mixture is just another representation of the original model, which can be denoted
\begin{equation}
	\label{eq:merge}
	f(\mathbf{x}) = \sum_{j=1}^{H}\pi_j^{*}f_j^* (\mathbf{x}) = \sum_{g=1}^G \pi_g \phi(\mathbf{x} \mid \boldsymbol{\mu}_g, \boldsymbol{\Sigma}_g),
\end{equation}  
where each $\pi_j^*$ is equivalent to one of or the sum of some of the mixing proportions $\pi_1,\ldots,\pi_G$, and each $f_j^* (\mathbf{x})$ is given by one of or a mixture of some of the component densities $\phi(\mathbf{x}~|~\boldsymbol{\mu}_1, \boldsymbol{\Sigma}_1),\ldots,\phi(\mathbf{x}~|~\boldsymbol{\mu}_G, \boldsymbol{\Sigma}_G)$. 
For example, suppose that $G=3$ and $H=2$; then, $f_1^* (\mathbf{x})$ could be a convex combination of the first two Gaussian components, in which case $f_2^* (\mathbf{x})$ would equal the third one. This gives us a straightforward method of thinking about merged components but does not tell us which components should be merged with one another, e.g., when we considered the example with $G=3$ and $H=2$, we looked at only one of three possible merging outcomes.

We develop a mixture model component merging criterion based on the ARI, which is the Rand index \citep{rand71} corrected for chance agreement. The Rand index compares two data partitions based on pairwise agreement and is given by the number of pairwise agreements divided by the total number of pairs. The correction leading to the ARI is performed to account for the fact that some cases will be correctly classified by chance if classification is performed randomly. An ARI value of 1 corresponds to perfect class agreement and the expected value of the ARI under random classification is 0. Some properties of the ARI are discussed by \cite{steinley04}.

Herein, we use the ARI to merge components based on a `reference model'. The purpose of merging components is that models in Occam's window have the same number of components as the reference model so that we can use them all if desired. Two cases are considered. 
\begin{itemize}
\item In Case~I, the model with the smallest BIC is the reference model. Merging is performed for models in Occam's window with more components than the reference model; however, models in Occam's window with fewer components than the reference model are discarded. 
\item In Case~II, the model with the fewest components is the reference model and merging is performed on other models in Occam's window, as needed, to give the same number of components.
\end{itemize} 
Of course, Case I and Case II are equivalent when the model with the smallest BIC also has the fewest components. Note that in Case~I we are effectively assuming that the BIC will only overestimate the number of components; however, we take a more flexible position in this regard in Case~II.

Our merging process is best illustrated with an example. Suppose that one model inside Occam's window has seven components and that the reference model is a four-component model. The partition corresponding to the reference model is used as the underlying `true' classification in the merging process and is called the `reference partition'. In what follows here, to avoid confusion, we use $\{1,2,\ldots ,7\}$ to denote components of the seven-component model and $\{a,b,c,d\}$ to denote components in the reference model. The following steps illustrate how our merging strategy works.
\begin{enumerate}
	\item A combination matrix $\mathbf{A}$ of size $35\times 4$ is generated. Note that $\binom{7}{4}=35$, where each row represents a partial clustering, e.g., if the 22nd row of $\mathbf{A}$ is $\mathbf{a}_{22}=(2, 3, 4, 6)$, then component 2 goes into new component $a$, component 3 goes into new component $b$, component~4 goes into new component $c$, and component~6 goes into new component~$d$.
	\item For each row in $\mata$, we must deal with the remaining components, e.g., $\{1, 5, 7\}$ in the case of row~22 with respect to the components $a, b, c, d$ introduced above. A permutation matrix $\mathbf{B}$ ($64 \times 3$) is created to include all possibilities. One sample row is $\mathbf{b}_{j}=(a, a, d)$, which puts component 1 into component $a$, component 5 into component $a$, and component~7 into component $d$, so the model after merging has the new components $\{a,b,c,d\}=\{1\cup2\cup5,3,4,6\cup7\}$.
	\item The ARIs between the reference partition and all partitions arising from the model after merging are calculated and recorded in the $35\times 64$ matrix~$\mathbf{C}$. For example, in step 1, $\mathbf{a}_{22}=(2, 3, 4, 6)$ is the $22\text{nd}$ row in matrix $\mathbf{A}$, in step 2, $\mathbf{b}_{j}=(a, a, d)$ is the $4\text{th}$ row in matrix $\mathbf{B}$, and the ARI value between reference partition and this model after merging is stored at the $22\text{th}$ row and the $4\text{th}$ column of the matrix $\mathbf{C}$. 
\end{enumerate} 
Once we have every element of the matrix $\mathbf{C}$, the best merging combination is chosen to correspond to the largest ARI value, i.e., the largest element in $\mathbf{C}$. 
 
\subsection{Averaging \textit{A Posteriori} Probabilities}\label{sec:aap}
In model-based clustering applications, we use $z_{ig}$ to denote component membership, where $z_{ig}=1$ if observation~$i$ is in component~$g$ and $z_{ig}=0$ otherwise. After parameters have been estimated, e.g., via the EM algorithm, the \textit{a~posteriori} probabilities (or, equivalently, expected values)
\begin{equation}\label{eqn:z}
\hat{z}_{ig}\colonequals\frac{\hat{\pi}_g\phi(\vecx_i\mid\hat{\vecmu}_g,\hat{\matsig}_g)}{\sum_{h=1}^G\hat{\pi}_h\phi(\vecx_i\mid\hat{\vecmu}_h,\hat{\matsig}_h)}
\end{equation}
are used to produce the predicted classifications. Typically, the $\hat{z}_{ig}$ are hardened to either 0 or 1 based on maximum \textit{a posteriori} (MAP) probabilities, i.e., $\text{MAP}(\hat{z}_{ig})=1$ if $\max_h\{\hat{z}_{ih}\}$ occurs in component $h=g$, and $\text{MAP}(\hat{z}_{ig})=0$ otherwise. See \cite{basford85} for a discussion of allocation of observations into clusters in the model-based clustering context. Within our averaging approaches, if components of a model in Occam's window need to be merged, then, using the notation in \eqref{eq:merge}, the post-merging $\hat{z}_{ij}^*$ will correspond to one of or the sum of pre-merging $\hat{z}_{ig}$, for $j=1,\ldots,H$ and $g=1,\ldots,G$. For example if $f_1^*(\vecx_i)$ is a convex combination of what were the first two components, then we simply have $\hat{z}_{i1}^*=\hat{z}_{i1}+\hat{z}_{i2}$.

After any necessary merging has been carried out for models within Occam's window, and models are dropped if necessary (i.e., if in Case~I there are fewer components than the reference model), we compute a weighted average of the \textit{a posteriori} probabilities. For each~$i$, we average the \textit{a posteriori} probabilities from the models $\mathcal{M}_1, \mathcal{M}_2,\ldots,\mathcal{M}_K$ using the weights $\text{Pr} ( \mathcal{M}_i \mid D)$, cf.\ \eqref{eq:postmodel2}. Averages of the \textit{a posteriori} probabilities can then be hardened to give predicted classifications.

\subsection{Model Averaging}\label{sec:ma}
In addition to averaging \textit{a posteriori} model probabilities, we also consider a direct averaging of models. This time, we only consider models within Occam's window that have the same number of components as the model with the smallest BIC. 
We compute a weighted average of the parameter estimates for each parameter in these models $\mathcal{M}_i$ using the weights $\text{Pr}(\mathcal{M}_i~|~D)$, cf.\ \eqref{eq:postmodel2}. Expected values $\hat{z}_{ig}$ are then computed based on the averaged parameter estimates and the predicted classifications are based on the associated maximum \textit{a~posteriori} probabilities.

This model averaging approach is quite different from averaging \textit{a posteriori} probabilities (Section~\ref{sec:aap}). Rather than focusing on a weighted average of \textit{a posteriori} probabilities, possibly after merging, here we are concerned with producing classifications from one interpretable model. That is, the classifications from our model averaging approach will come from one parameterized, interpretable mixture model, e.g., if we average $G$-component models from the GPCM family, the result of model averaging is a $G$-component Gaussian mixture model. 

Note that for all of our averaging approaches, we need to be careful to correctly match components across models, e.g., component~1 in the best model might be equivalent to component~2 in the second best model. This issue is related to the problem of label switching in mixture models \citep[cf.][]{stephens00}. In both averaging strategies, we match components based on minimum distance between estimated component means.

\subsection{Comments}
Although more generally applicable, we develop and demonstrate our averaging approaches for Gaussian mixture models. Naturally, questions arise around what sort of situations might benefit from these approaches and what other methods are appropriate for comparison. Recall that the objective of our averaging approaches is to improve on the clustering performance of a single, i.e., the best, (Gaussian) mixture model. In the case of model averaging (Section~\ref{sec:ma}), the result is itself a Gaussian mixture model. Accordingly, model averaging is expected to work well for clustering in the same kind of situations where Gaussian mixture models would be expected to work well. When averaging \textit{a~posteriori} probabilities in Case~I (Section~\ref{sec:aap}), the best model is the reference model, models with fewer components are discarded, and merging is used for models with more components than the reference model before averaging is carried out. Because the best model is the reference model, averaging \textit{a~posteriori} probabilities in Case~I tends to produce similar clustering results to model averaging, and this is borne out in our real and simulated data analyses (Section~\ref{sec:data}).  

When averaging \textit{a~posteriori} probabilities in Case~II (Section~\ref{sec:aap}), the model with the fewest components is the reference model and all models in Occam's window are used for averaging (after merging is carried out where needed). Unless the best model is also the model with the fewest components, averaging \textit{a~posteriori} probabilities in Case~II produces clusters that may be quite different from those produced by model averaging or averaging \textit{a~posteriori} probabilities in Case~I. There are two reasons for this: the number of components is different and more models are used in the averaging. That is not to say that the clustering results arising from averaging \textit{a~posteriori} probabilities in Case~II will be very different to those produced by a single Gaussian mixture model --- after all, we know there is a well-fitting (as determined by being inside Occam's window) single Gaussian mixture with the same number of components --- but rather that clustering results arising from averaging \textit{a~posteriori} probabilities in Case~II might be quite different to those produced by the best model or one of our other averaging approaches. This is also borne out in our real and simulated data analyses (Section~\ref{sec:data}).

It is possible that averaging \textit{a~posteriori} probabilities in Case~II might successfully capture clusters that are not well-described by a Gaussian distribution; however, it can only do this to the extent that a model within Occam's window has selected the appropriate number of components (e.g., Section~\ref{sec:sim}, Scenario~III). Because of this limitation, we do not proffer averaging \textit{a~posteriori} probabilities in Case~II, or either of our other averaging approaches, as techniques for finding clusters that cannot be captured using a single Gaussian distribution. In such cases, a non-Gaussian mixture model could be used, \textit{a~posteriori} merging of components (cf.\ Section~\ref{sec:apmc}) could be applied to the best model or to the model that results from model averaging (e.g., Section~\ref{sec:sim}, Scenario~III), or the approaches used herein could be applied to average non-Gaussian mixture models.

\section{Data Analyses}\label{sec:data}
\subsection{Performance Assessment}
All of our examples are treated as genuine clustering examples, i.e., no prior knowledge of the labels or the number of components is used. However, the true labels are known in each case and we can use them to evaluate classification performance. There are many indices that can be used to compare agreement between true and predicted classes. In a comprehensive comparison, \cite{milligan86} recommended the ARI, which has become the index of choice in model-based clustering applications. As mentioned in Section~\ref{sec:method}, \cite{steinley04} discusses some properties of the ARI. We use the ARI to evaluate classification performance herein. 

\subsection{Real Data Analyses}

\subsubsection{Italian wine}
\cite{forina86} present data on chemical and physical properties of 178 samples of three varieties (Barolo, Barbera, and Grignolino) of wine from the Piedmont region of Italy. A data set containing 27 physical and chemical properties is available in the {\tt pgmm} package \citep{mcnicholas11b} for {\sf R}. We fit the GPCM models to these data using {\tt mixture}, under the default settings, for $G = 1, 2, \dots, 9$. The VVI model with $G = 3$ components is the best model in terms of the BIC ($12103.74$), but the EVI model has an almost identical BIC value ($12103.81$). This is an example where choosing the `best' model in terms of the BIC and ignoring all other models is clearly questionable because the BIC values are so close. Although the BIC values are extremely close, the classifications associated with the respective models are different, i.e., the respective MAP classifications disagree on four of the wine samples.

We apply our averaging procedures to these data. Only the aforementioned two models lie in Occam's window and, because both have $G = 3$ components, merging components is not required.  The results (\tablename~\ref{winetopm}) indicate that averaging \textit{a~posteriori} probabilities leads to a very slight improvement in the ARI over the best model (going from $0.8951$ to $0.8962$), while model averaging gives a slightly inferior ARI ($0.8456$) when compared to the best model. 	
\begin{table}[!ht]{\small
	\caption{A summary of the models in Occam's window along with ARI values for the true labels versus predicted classifications from the best model, from averaging \textit{a~posteriori} probabilities (AAP), and from model averaging (MA), respectively, for the Italian wine data.}
	\label{winetopm}
	\centering
	\begin{tabular*}{1.0\textwidth}{@{\extracolsep{\fill}}lcccccccc}
	\hline
		 \multicolumn{3}{c}{Occam's Window} &  & $\text{Pr}(\mathcal{M}_i \mid D)$ & &  \multicolumn{3}{c}{ARI Values}\\
		 \cline{1-3}\cline{7-9}
		    Model & BIC & $G$ &&  &&Best& AAP & MA\\
	\hline
   VVI &   $12103.74$ & 3 & &  0.5088 && \multirow{2}{*}{0.8951}&\multirow{2}{*}{0.8962}&\multirow{2}{*}{0.8456}\\
   EVI &   $12103.81$ & 3 &         &  0.4912 && \\
    \hline		                    
	\end{tabular*}}
\end{table}

\subsubsection{Female voles data}
\cite{flury97} discusses seven measurements on female voles from two species, \textit{Microtus californicus} and \textit{Microtus ochrogaster}. The data are available within the {\sf R} package {\tt Flury} \citep{flury12}. The GPCM models are fitted to these data using {\tt mixture}, for $G = 1, 2, \dots, 9$. The BIC selected a VEE model with $G = 3$ components and ARI = $0.6577$. We apply our averaging procedures to these data. There are three models within Occam's window, i.e., the best model and two models with $G=2$ components. When averaging \textit{a posteriori} probabilities, we merge components of the best model to give a two-component model (i.e., Case~II). Averaging \textit{a posteriori} probabilities leads to much better classification performance (ARI = $0.9081$) when compared with the best model ($\text{ARI} = 0.6577$), cf.\ \tablename~\ref{fvolesapp}. Because we only consider model averaging when there are models within Occam's window with the same number of components as the best model, it is not applicable here. 
\begin{table}[!ht]{\small
	\caption{A summary of the models in Occam's window along with ARI values for the true classifications versus classifications from the best model and from averaging \textit{a posteriori} probabilities (AAP), respectively, for the female voles data.}
	\label{fvolesapp}
	\centering
	\begin{tabular*}{1.0\textwidth}{@{\extracolsep{\fill}}lccccccc}
	\hline
		 \multicolumn{3}{c}{Occam's Window} &  & $\text{Pr}(\mathcal{M}_i \mid D)$ & &  \multicolumn{2}{c}{ARI Values}\\
		 \cline{1-3}\cline{7-8}
		    Model & BIC & $G$ && &&Best& AAP \\
	\hline
   VEE &   $1309.37$ & 3 & &  0.9540 && \multirow{3}{*}{0.6577}&\multirow{3}{*}{0.9081}\\
   EEE &   $1316.70$ & 2 &         &  0.0244 && \\
   VEE&   $1316.95$ & 2 &         &  0.0216 && \\
    \hline		                    
	\end{tabular*}}
\end{table}

\subsubsection{Swiss bank notes}\label{sec:swiss}
The Swiss bank note data are available in the {\tt mclust} package \citep{fraley13}, and comprise six measurements on 100 genuine and 100 counterfeit old Swiss 1000-franc bank notes. The GPCM models are fitted to these data using {\tt mixture}, for $G = 1, 2, \dots, 9$. The BIC selects an EEE model with $G = 4$ components and $\text{ARI}=0.6790$. This model uses two components for each of the true labels (i.e., genuine and counterfeit). 
We apply our averaging procedures to these data. This time, three models lie in Occam's window: the first (i.e., the best) and third models have $G=4$ components and the second model has $G=3$ components. 

When averaging \textit{a posteriori} probabilities, we must either ignore the three-component model (Case~I) or merge components within the four-component models to give three-component models (Case~II). 
The results (\tablename~\ref{bankaap}) indicate that averaging \textit{a~posteriori} probabilities under Case~I leads to the same classification performance as given by the best model (i.e., ARI = $0.6790$); however, averaging \textit{a posteriori} probabilities under Case~II leads to a great improvement in classification performance ($\text{ARI} = 0.9068$). When carrying out model averaging, we only consider the two models with $G = 4$ components, i.e., the first and third models, and the results (\tablename~\ref{bankaap}) show that model averaging has led to an improved classification performance (ARI = $0.7602$) when compared with the best model (ARI = $0.6790$). However, the classification improvement using model averaging is notably less than that from averaging \textit{a~posteriori} probabilities in Case~II. 
\begin{table}[!ht]
	{\small\caption{A summary of the models in Occam's window along with ARI values for the true labels versus predicted classifications from the best model, from averaging \textit{a posteriori} probabilities (AAP), and from model averaging (MA), respectively, for the bank note data.}
	\label{bankaap}
	\centering{
	\begin{tabular*}{1.0\textwidth}{@{\extracolsep{\fill}}lcccccccccc}
	\hline
		 \multicolumn{3}{c}{Occam's Window} &  &   \multicolumn{2}{c}{$\text{Pr}(\mathcal{M}_i \mid D)$} & &  \multicolumn{4}{c}{ARI Values}\\
		 \cline{1-3}\cline{5-6}\cline{8-11}
		    Model & BIC & $G$ && Case~I &Case~II &&Best& \multicolumn{2}{c}{AAP}&MA\\
		    \cline{9-10}
		    & & &&/ MA & && & Case I & Case II\\
	\hline
   EEE &   $2651.92$ & 4 & &  0.9819 & 0.6736&& \multirow{3}{*}{0.6790}&\multirow{3}{*}{0.6790}&\multirow{3}{*}{0.9068}&\multirow{3}{*}{0.7602}\\
   VEE &   $2653.44$ & 3 &&   &0.3141& \\
   VEE &   $2659.91$ & 4 &  & 0.0181& 0.0124 && \\
    \hline		                    
	\end{tabular*}}}
\end{table}

\subsubsection{Iris data}
\cite{fisher36} discussed four measurements, in centimetres, of the length and width of sepals and petals from three species of irides (\textit{Iris setosa}, \textit{Iris virginica}, and \textit{Iris versicolor}), originally collected by \cite{anderson35}. The data consist of 50 samples from each of three species and is available in the {\tt datasets} package for {\sf R}. The GPCM models are fitted to these data using {\tt mixture}, for $G = 1,2,\dots,9$. The BIC selected an VEV model with $G = 3$ components and $\text{ARI} = 0.9222$. 
Three models lie in Occam's window and they do not have the same number of components. Similar to our analysis of the Swiss bank notes data (Section~\ref{sec:swiss}), we must either ignore one model (Case~I) or merge components within the three-component models to give two-component models (Case~II). Averaging \textit{a posteriori} probabilities under Case~I leads to the same classifications as given by the best model (i.e., $\text{ARI} = 0.9222$), while Case~II gives an inferior ARI ($0.5681$) when compared to the best model (cf.\ \tablename~\ref{irisapp}). Because there are three species of iris in these data, it is not surprising that Case~II does not yield a favourable result here. Model averaging is also performed and gives a slightly inferior ARI ($0.9039$) when compared to the best model (\tablename~\ref{irisapp}).
\begin{table}[!ht]
{\small\caption{A summary of the models in Occam's window along with ARI values for the true labels versus predicted classifications from the best model, from averaging \textit{a posteriori} probabilities (AAP), and from model averaging (MA), respectively, for the iris data.}	\label{irisapp}
	\centering
	\begin{tabular*}{1.0\textwidth}{@{\extracolsep{\fill}}lcccccccccc}
	\hline
		 \multicolumn{3}{c}{Occam's Window} &  &   \multicolumn{2}{c}{$\text{Pr}(\mathcal{M}_i \mid D)$} & &  \multicolumn{4}{c}{ARI Values}\\
		 \cline{1-3}\cline{5-6}\cline{8-11}
		    Model & BIC & $G$ && Case~I &Case~II &&Best& \multicolumn{2}{c}{AAP}&MA\\
		    \cline{9-10}
		    & & &&/ MA & && & Case~I & Case~II\\
	\hline
   VEV &   $789.37$ & 3 & &  0.9833 & 0.6531&& \multirow{3}{*}{0.9222}&\multirow{3}{*}{0.9222}&\multirow{3}{*}{0.5681}&\multirow{3}{*}{0.9039}\\
   VVV &   $790.70$ & 2 &&                        &0.3358 && \\
   VVV &   $797.52$ & 3 &  & 0.0167& 0.0111 && \\
    \hline		                    
	\end{tabular*}}
\end{table}

\subsubsection{Coffee data}
The coffee data comprise 43 observations on twelve chemical properties of coffee samples collected from two species (Arabica and Robusta) from 29 countries around the world. These data are available in the {\tt pgmm} package for~{\sf R}. The {\tt mixture} package is used to fit the GPCM family to the twelve chemical properties for $G=1,\ldots,9$. The VEI model with $G = 2$ was chosen by the BIC, and there is one other model within Occam's window, i.e., a $G=2$ component EEI model. Performing model averaging via \textit{a posteriori} probabilities or model averaging leads to perfect classification (ARI = $1$; \tablename~\ref{coffeeappma}), as does the best model.
\begin{table}[!ht]
{\small\caption{A summary of the models in Occam's window along with ARI values for the true labels versus predicted classifications from the best model, from averaging \textit{a posteriori} probabilities (AAP), and from model averaging (MA), respectively, for the coffee data.}	\label{coffeeappma}
	\centering
	\begin{tabular*}{1.0\textwidth}{@{\extracolsep{\fill}}lcccccccc}
	\hline
		 \multicolumn{3}{c}{Occam's Window} &  & $\text{Pr}(\mathcal{M}_i \mid D)$ & &  \multicolumn{3}{c}{ARI Values}\\
		 \cline{1-3}\cline{7-9}
		    Model & BIC & $G$ &&  &&Best& AAP & MA\\
	\hline
   VEI &   $1334.22$ & 2 & &  0.6843 && \multirow{2}{*}{1}&\multirow{2}{*}{1}&\multirow{2}{*}{1}\\
   EEI &   $1335.72$ & 2 &         &  0.3157 && \\
    \hline		                    
	\end{tabular*}}
\end{table}

\subsubsection{Examples with only one model in Occam's window: breast cancer data and olive oil data}
In a Wisconsin breast cancer study, several features are recorded for 681 cases of potentially cancerous tumours, of which 238 are actually malignant. \cite{mangasarian95} use these data to establish whether fine needle aspiration can determine tumour status. These data are available in the {\tt faraway} package \citep{faraway11} for {\sf R}.

The olive oil data, available in the {\tt pgmm} package for {\sf R}, comprise 572 observations on the percent composition of olive oil with respect to eight fatty acids found by lipid fractionation from nine areas across three regions (Southern Italy, Sardinia, and Northern Italy) in Italy. 

Fitting the GPCM models to each of these data sets using {\tt mixture}, for $G = 1, 2, \dots, 9$, results in only one model lying within Occam's window. For the olive oil data, the BIC chooses an EVE model with $G = 9$ components. For the breast cancer data, the BIC chooses an EVE model with $G = 6$ components. We note that $G=9$ is the upper end of the range of values that was used for $G$ (i.e., $G=1,\ldots,9$); accordingly, we repeated the analysis of these data using the range $G=1,\ldots,20$. The same result was obtained, i.e., the EVE model with $G = 9$ components was the only model inside Occam's window. Of course, when there is only one model in Occam's window, averaging is equivalent to reporting classifications from the model with the best BIC.

\subsection{Simulated Data}\label{sec:sim}
Three simulation scenarios are considered. In each case, data are generated via the {\tt{genRandomClust()}} function from the {\sf R} package {\tt clusterGeneration} \citep{qui12}. The {\tt genRandomClust()} function generates random clusters based on the method proposed by \citet{qiu06}. 
\begin{figure}[!ht]
	\centering
	\includegraphics[width=0.75\textwidth]{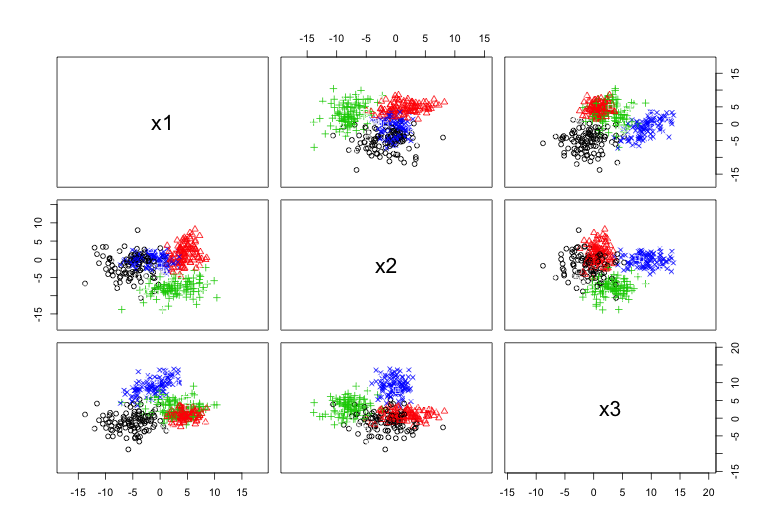}
	\vspace{-0.2in}
	\caption{A pairs plot for one of the 30 data sets from Scenario~I.}
	\label{simulatepairs}
\end{figure}

In Scenario~I, we generate 30 data sets with $p=3$ variables, $n=480$ observations, and substantially overlapping clusters by setting {\tt numClust=4}, {\tt sepVal=0.03}, {\tt numNonNoisy=3}, and {\tt clustszind=2} (e.g., \figurename~\ref{simulatepairs}). This is a very difficult clustering problem and perfect classification results are not expected. We use {\tt mixture} to fit each of the 30 data sets for $G = 1, 2, \dots, 9$. 
%
Over the 30 runs, the ARI values achieved by the best model, averaging \textit{a~posteriori} probabilities (Cases~I and~II), and model averaging were all similar (Table ~\ref{wontie}). However, the performance of averaging \textit{a~posteriori} probabilities in Case~II might be considered slightly preferable to the others because it has the largest mean ARI as well as the smallest standard deviation. 
\begin{table}[!ht]{\small
\caption{The summary for the best model, averaging \textit{a posteriori} probabilities (AAP), and model averaging (MA) among 30 runs from Scenario~I.}
	\label{wontie}
	\centering
	\begin{tabular*}{1.0\textwidth}{@{\extracolsep{\fill}}lcccccc}
	\hline
		ARI & Best &  &   \multicolumn{2}{c}{AAP} & &  MA \\
		 \cline{4-5}
		 &                      &  & Case~I & Case~II& &\\
	\hline
          Average&0.7345&&0.7373&0.7528&&0.7442\\
          Std.\ Deviation &0.0624&&0.0631&0.0367&&0.0604\\
          \hline		                    
	\end{tabular*}}
\end{table}

In Scenario~II, we generate $p=3$ variables for $n=644$ observations and we add one noise variable (by setting {\tt numNoisy=1}). We used {\tt mixture} to fit the GPCM models to these data for $G = 1, 2, \dots, 9$. The BIC selects a VEE model with $G=4$ and $\text{ARI}=0.9061$, and there are four models within Occam's window (\tablename~\ref{simulatenoise}). One of the models (EEI, $G=6$) has more components than the best model and so merging is required before averaging \textit{a~posteriori} probabilities. Both model averaging ($\text{ARI}=0.9196$) and averaging \textit{a~posteriori} probabilities ($\text{ARI}=0.9092$) lead to a slight improvement in classification performance, with the former being slightly better. 
\begin{table}[!ht]{\small
	\caption{A summary of the models in Occam's window along with ARI values for the true labels versus predicted classifications from the best model, from averaging \textit{a posteriori} probabilities (AAP), and from model averaging (MA), respectively, for the simulated data from Scenario~II.}
	\label{simulatenoise}
	\centering
	\begin{tabular*}{1.0\textwidth}{@{\extracolsep{\fill}}lccccccccc}
	\hline
		 \multicolumn{3}{c}{Occam's Window} && \multicolumn{2}{c}{$\text{Pr}(\mathcal{M}_i \mid D)$} && \multicolumn{3}{c}{ARI Values}\\
		 \cline{1-3}\cline{5-6}\cline{8-10}
		    Model & BIC & $G$ &&  AAP& MA &&Best& AAP & MA\\
	\hline
   VEE &   $11981.63$ & 4 & &  0.4410 &0.4504 && \multirow{4}{*}{0.9061}&\multirow{4}{*}{0.9092}&\multirow{4}{*}{0.9196}\\
   EEE &   $11982.15$ & 4 &         &  0.3401 & 0.3473&& \\
   EVE &   $11983.23$ & 4 &         &  0.1990 & 0.2024&& \\
    EEI &   $11987.83$ & 6 &         &  0.0199 & && \\
    \hline		                    
	\end{tabular*}}
\end{table}

\begin{figure}[!ht]
	\centering
	\includegraphics[width=0.7\textwidth]{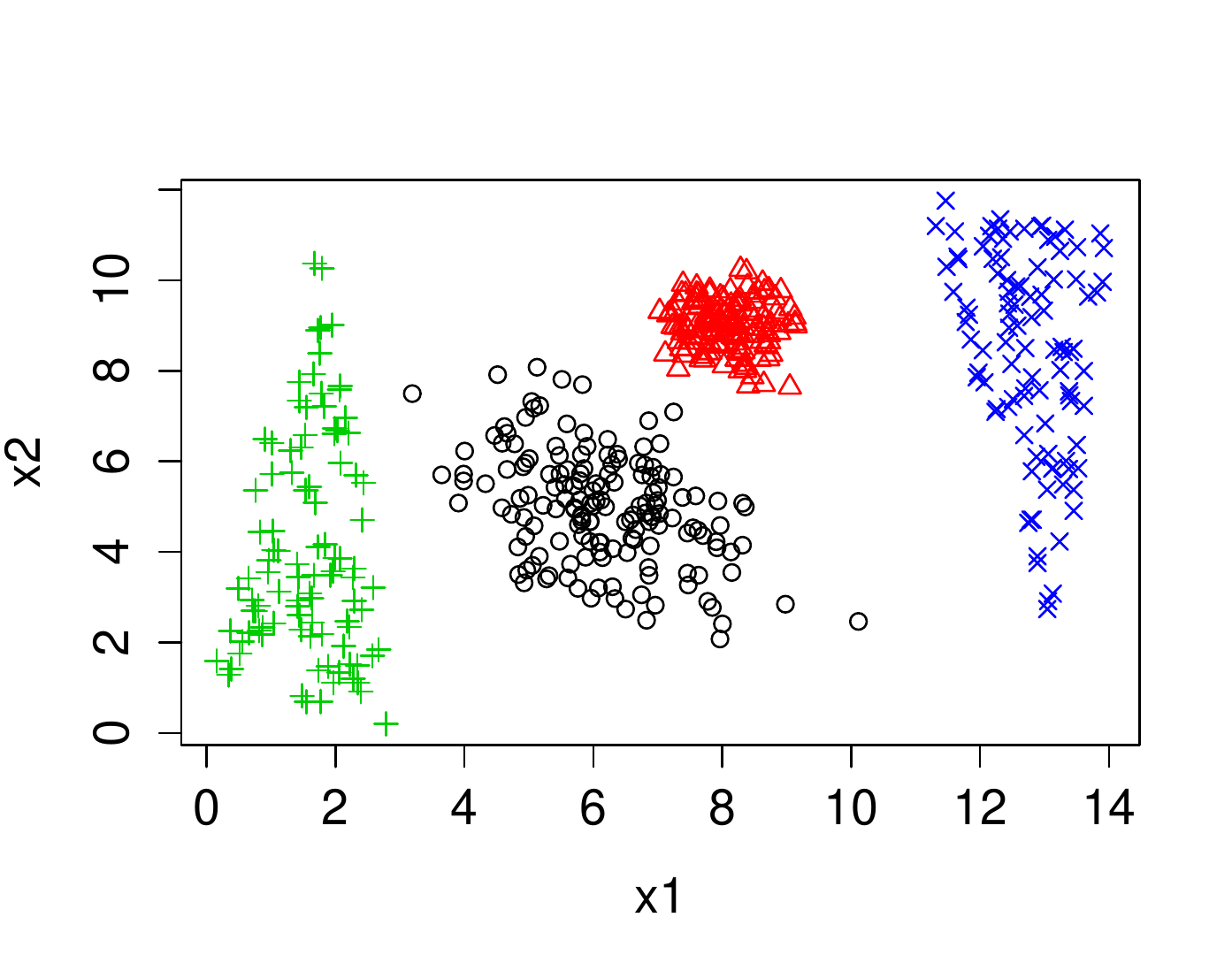}
	\vspace{-0.2in}
	\caption{Scatterplot of the simulated data from Scenario~III.}
	\label{simulatescatter}
\end{figure}

In Scenario~III, we generated four components. Two are simulated from bivariate Gaussian distributions, and the other two are uniformly generated from triangles via the rejection method (\figurename~\ref{simulatescatter}). There are $500$ points, $100$ in each of the triangles and $150$ in each of the Gaussian components. These data were generated to illustrate our merging approaches when the model generating the data is very clearly not within the family of models being used for analysis. 

\begin{figure}[!ht]
		\ \includegraphics[width=0.5\textwidth]{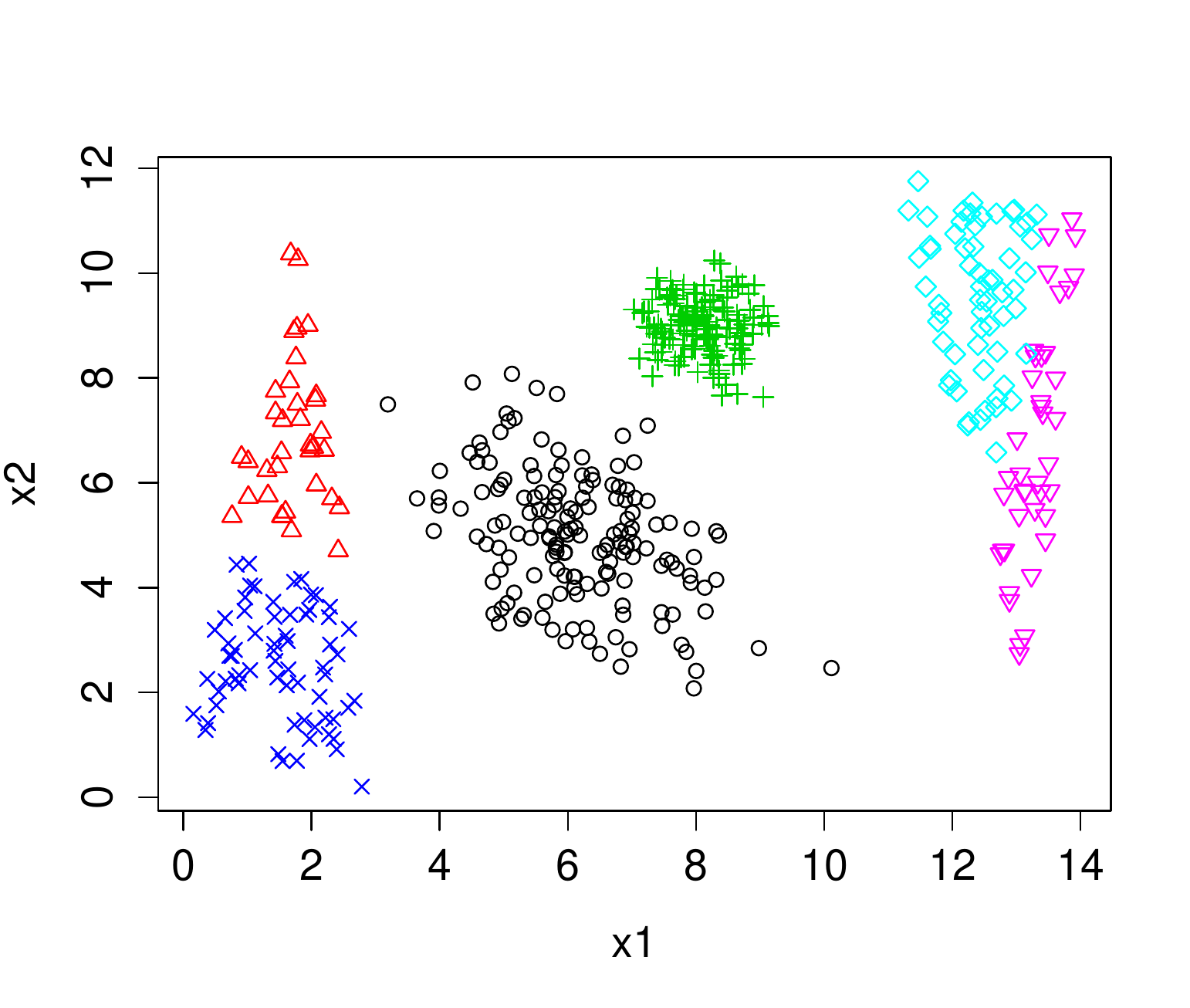} \ \ \
		\includegraphics[width=0.5\textwidth]{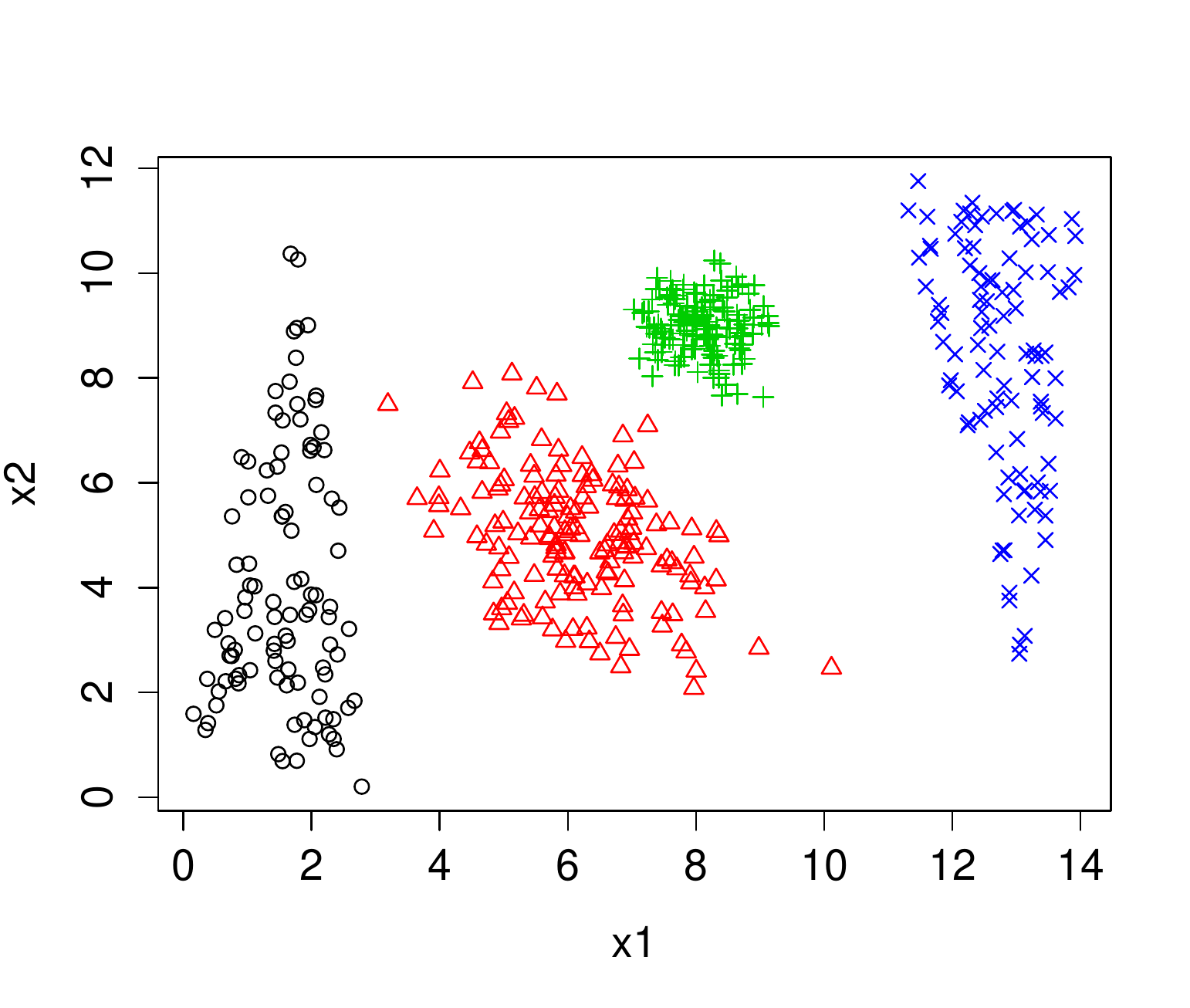}
		\vspace{-0.2in}
	\caption{The best model (VVV, $G=6$, left-hand side) and one other model from within Occam's window (VVV, $G=4$, right-hand side) for Scenario~III.}\label{simu4BIC}
\end{figure}

We fit the GPCM models to these data using {\tt mixture}, for $G=1,\ldots,9$. A VVV model with $G=6$ components and $\text{ARI}=0.9061$ was selected by the BIC (\figurename~\ref{simu4BIC}), and there are four other models in Occam's window (with $G=4,5,6,$ and~$7$ components, respectively). Averaging \textit{a posteriori} probabilities in Case~I leads to some improvement in ARI ($0.9344$), and in Case~II leads to perfect classification (\tablename~\ref{simulationaap}, \figurename~\ref{simu4BIC}). Model averaging gives the same classification performance as the best model (\tablename~\ref{simulationma}); notably, if we merge the components of the resulting model, perfect classification results are obtained (\figurename~\ref{simu4BIC}). This is a nice illustration of how model averaging can be followed by component merging to model non-Gaussian clusters.
\begin{table}[!ht]{\small
	\caption{A summary of the models in Occam's window along with ARI values for the true labels versus predicted classifications from the best model and from averaging \textit{a posteriori} probabilities (AAP), respectively, for the simulated data from Scenario~III.}
	\label{simulationaap}
	\centering
	\begin{tabular*}{1.0\textwidth}{@{\extracolsep{\fill}}lccccccccc}
	\hline
		 \multicolumn{3}{c}{Occam's Window} &  &   \multicolumn{2}{c}{$\text{Pr}(\mathcal{M}_i \mid D)$} & &  \multicolumn{3}{c}{ARI Values}\\
		 \cline{1-3}\cline{5-6}\cline{8-10}
		    Model & BIC & $G$ && Case~I &Case~II &&Best& \multicolumn{2}{c}{AAP}\\
		    \cline{9-10}
		    & & && & && & Case~I & Case~II\\
	\hline
   VVV &   $4168.84$ & 6 & &  0.6820 & 0.5471&& \multirow{5}{*}{0.9061}&\multirow{5}{*}{0.9344}&\multirow{5}{*}{1}\\
   VVI &   $4170.39$ & 6 &&     0.3145 &0.2523&& \\
   VVV &   $4171.37$ & 5 &  & & 0.1547 && \\
   VVV& $ 4173.92$&4&&   &0.0431&&\\
   VVI &   $4179.37$&7&&   0.0035  &0.0028&&\\ 
    \hline		                    
	\end{tabular*}}
\end{table}
\begin{table}[!ht]{\small
	\caption{A summary of the models in Occam's window along with ARI values for the true labels versus predicted classifications from the best model and from model averaging (MA), respectively, for the simulated data from Scenario~III.}
	\label{simulationma}
	\centering
	\begin{tabular*}{1.0\textwidth}{@{\extracolsep{\fill}}lccccccc}
	\hline
		 \multicolumn{3}{c}{Occam's Window \& $G=6$} &  & $\text{Pr}(\mathcal{M}_i \mid D)$ & &  \multicolumn{2}{c}{ARI Values}\\
		 \cline{1-3}\cline{7-8}
		    Model & BIC & $G$ &&  &&Best& MA\\
	\hline
   VVV &   $4168.84$ & 6 & &  0.6844 && \multirow{2}{*}{0.9061}&\multirow{2}{*}{0.9061}\\
   VVI &   $4170.39$ & 6 &   &  0.3156 && \\
    \hline		                    
	\end{tabular*}}
\end{table}

\subsection{Comparison with Strehl's Consensus Clustering Algorithms}
In this section, our methodology is compared with the consensus clustering algorithms  proposed by \cite{strehl02}. Consensus clustering is defended as forming a single consolidated clustering through combining multiple clusterings of a set of objects without considering the features or algorithms that determined these clusterings. \cite{strehl02} introduced three efficient heuristics to solve the cluster ensemble problem: CSPA, HGPA, and MCLA (cf.\ Section~\ref{sec:intro}). These three algorithms transform the given multiple clusterings into a suitable graph representation. Note that MCLA uses hyperedge collapsing operations to determine soft cluster membership values for each object and is a more efficient algorithm than CSPA or HGPA. MATLAB \citep{matlab10} code for these three algorithms is available at {\tt http://strehl.com}. To facilitate comparison of Strehl's algorithms to our methodology, the partitions corresponding to the models in Occam's window are taken to be the multiple clusterings that need to be consolidated. 

We compare our methods to those using Strehl's consensus clustering algorithms for each of the five real data sets already considered that had multiple models inside Occam's window (\tablename~\ref{comparison}). For the coffee data, there is a three-way tie between averaging \textit{a~posteriori} probabilities, model averaging, and MCLA. In the other four cases, one of our averaging \textit{a~posteriori} probabilities approaches outperforms all other approaches. In the four cases were model averaging was applied, it gave the same classification performance as MCLA in one case and outperformed all consensus clustering approaches in the other three cases. Looking at the three consensus clustering algorithms, we note that MCLA always outperforms both CSPA and HGPA. The very poor performance of both CSPA and HGPA on the coffee data is also noteworthy, especially when one considers that all other methods give perfect classification results.
\begin{table}[!ht]{\small
	\caption{A summary of ARI values for our methodology and Strehl's algorithms on five real data sets. The highest ARI for each data set is highlighted.}
	\label{comparison}
	\centering
	\begin{tabular*}{1.0\textwidth}{@{\extracolsep{\fill}}lcccccc}
	\hline
		 \multirow{2}{*}{} & \multicolumn{2}{c}{AAP}  & \multirow{2}{*}{MA} &\multirow{2}{*}{CSPA} &\multirow{2}{*}{HGPA}  &\multirow{2}{*}{MCLA}\\
		 \cline{2-3}
		    & Case~I & Case~II &&  &&\\	  
		  \hline  
		    Wine & \textbf{0.8962} &&0.8456&0.7590&0.1522&0.8625\\   
		    Female voles&&\textbf{0.9081}&&0.4488&0.3802&0.7060\\ 
		    Bank note&0.6790&\textbf{0.9068}&0.7602&0.4868&0.0315&0.6790\\
		    Iris&\textbf{0.9222}&0.5681&0.9039&0.8508&0.2889&0.9039\\
		    Coffee&\textbf{1}&&\textbf{1}&0.1100&0.0792&\textbf{1}\\
		  \hline            
	\end{tabular*}}
\end{table}

\section{Discussion}\label{sec:conc}

This paper proposes a departure from the `single best model' paradigm that has heretofore dominated the model-based clustering literature. Two averaging approaches are used and both focus on averaging models within Occam's window: one averages \textit{a posteriori} probabilities and the other averages the models' parameters to give a single interpretable model. Averaging \textit{a~posteriori} probabilities may require some components to be merged, and we introduced an ARI-based method to do this. This method requires a reference model, which is either taken to be the best model within Occam's window or the model within Occam's window with the fewest components.


While we present our approaches as either averaging \textit{a posteriori} probabilities or averaging models, one could also think of them as either using all models in Occam's window or using just the models with the same number of components as the best model. 
When using model averaging, we use the models with the same number of components as the best model. When averaging \textit{a posteriori} probabilities in Case~I, we use the models with at least as many components as the best model. When averaging \textit{a posteriori} probabilities in Case~II, we use all models to carry out the averaging. The performance of averaging \textit{a posteriori} probabilities will depend on the efficacy of the component merging approach used --- the method proposed herein (Section~\ref{sec:merging}) clearly worked very well on the real and simulated data considered in Section~\ref{sec:data}. The performance also depends on how good the choice of the number of components is; in fact, this is a central issue with all three of our approaches. Instead of choosing the number of components corresponding to the best model or choosing the fewest components, other approaches could be used to select the number of components. 

We used three simulation scenarios and seven real data sets to illustrate our approaches. On the seven real data sets, the performance of our averaging techniques was similar to, or slightly better than, that of the best model. In all but one case, averaging \textit{a~posteriori} probabilities in Case~II gave comparable or better performance to the best model and the other averaging approaches. The exception arose in the analysis of the iris data, where the number of components in the reference model was smaller than the true number of classes. Our approaches performed favourably when compared to the three consensus clustering algorithms introduced by \cite{strehl02}. 
Averaging \textit{a~posteriori} probabilities in Case~II has the advantage of using all models within Occam's window and perhaps has the greatest potential to notably improve classification performance; however, there is the risk that a model with too few components will result. Using model averaging or averaging \textit{a~posteriori} probabilities in Case~I avoids the risk of underestimating the number of components relative to the best model; however, the potential for improvement in classification performance may not be as great. 

One may argue that averaging is not needed, and that one should just report the set of competing models within Occam's window. Of course, there may be particular applications where this is preferable; however, end users generally want a single clustering result and our averaging approaches deliver this --- in fact, similar to model-based clustering in general, they deliver this in both hard and soft varieties. In terms of speed, it is noteworthy that the merging and averaging steps we use, i.e., the steps that we carry out above and beyond the usual model-based clustering procedure, are very fast and can be carried out for very large values of $n$, $G$, and $p$ (in fact, the value of latter is irrelevant in this respect). This means that our averaging approaches will not be a limiting factor in performing model-based clustering for data with larger values of $n$, $G$, and/or $p$, i.e., our averaging approaches can be applied to any model-based clustering technique that itself works well for the data at hand. In such cases, it might be worth considering the LASSO-penalized BIC \citep{bhattacharya14} as an alternative to the BIC, and this will be explored in future work.

The approaches introduced herein are perhaps the beginning of a corpus of work on averaging for families of mixture models in model-based clustering. We used the GPCM family of models to illustrate our approaches, but they could equally well be applied to non-Gaussian mixtures and to other families. We opted for $c=20$ in the definition of Occam's window \eqref{eq:owindow}, and the use of other values will be explored in future work. Alternatives to the BMA framework we consider could also be explored, leading to different weights and/or a different window. We focused on clustering applications herein, but our merging methods could also be applied, in an analogous fashion, for model-based classification \citep[e.g.,][]{dean06,mcnicholas10c,andrews11b} and model-based discriminant analysis \citep{hastie96}; this will also be a focus of future work.

\section*{Acknowledgements}
The authors are grateful to Professor Adrian Raftery and other members of the University of Washington Working Group on Model-Based Clustering for their comments and suggestions on an earlier version of this work. 
This work was supported by an Ontario Graduate Scholarship (Wei), a Discovery Grant from the Natural Sciences and Engineering Research Council of Canada (McNicholas), the University Research Chair in Computational Statistics at the University of Guelph (McNicholas), and an Early Researcher Award from the Ontario Ministry of Research and Innovation (McNicholas).  


\end{document}